\documentclass[sigconf]{acmart}

\AtBeginDocument{%
  \providecommand\BibTeX{{%
    \normalfont B\kern-0.5em{\scshape i\kern-0.25em b}\kern-0.8em\TeX}}}

\copyrightyear{2022} 
\acmYear{2022} 
\setcopyright{acmcopyright}\acmConference[ICCAD '22]{IEEE/ACM International Conference on Computer-Aided Design}{October 30-November 3, 2022}{San Diego, CA, USA}
\acmBooktitle{IEEE/ACM International Conference on Computer-Aided Design (ICCAD '22), October 30-November 3, 2022, San Diego, CA, USA}
\acmPrice{15.00}
\acmDOI{10.1145/3508352.3549429}
\acmISBN{978-1-4503-9217-4/22/10}

\usepackage{graphicx}
\usepackage{hyperref}
\usepackage{tikz}
\usepackage{comment}

\usepackage{amsmath,amssymb} 
\usepackage{color}
\usepackage{caption}
\usepackage{subcaption}
\usepackage{mathrsfs}
\usepackage{multirow}
\usepackage{booktabs}
\usepackage{caption}
\usepackage{subcaption}

\newcommand{\ourtit}{ObfuNAS}

\newcommand{\etal}{\textit{et al.}}

\begin{document}

\title{ObfuNAS: A Neural Architecture Search-based DNN Obfuscation Approach}

\author{Tong Zhou}

\affiliation{%
  \institution{Northeastern University}
  \city{Boston}
  \state{MA}
  \country{USA}
}
\email{zhou.tong1@northeastern.edu}

\author{Shaolei Ren}

\affiliation{%
  \institution{UC Riverside}
  \city{Riverside}
  \state{CA}
  \country{USA}}
\email{sren@ece.ucr.edu}

\author{Xiaolin Xu}

\affiliation{%
  \institution{Northeastern University}
  \city{Boston}
  \state{MA}
  \country{USA}
}
\email{x.xu@northeastern.edu}

\begin{abstract}
  Malicious architecture extraction has been emerging as a crucial concern for deep neural network (DNN) security. As a defense, architecture obfuscation is proposed to remap the victim DNN to a different architecture. Nonetheless, we observe that, with only extracting an obfuscated DNN architecture, the adversary can still retrain a substitute model with high performance (e.g., accuracy), rendering the obfuscation techniques ineffective. To mitigate this under-explored vulnerability, we propose \ourtit, which converts the DNN architecture obfuscation into a neural architecture search (NAS) problem. Using a combination of function-preserving obfuscation strategies, \ourtit\ ensures that the obfuscated DNN architecture can only achieve lower accuracy than the victim. We validate the performance of \ourtit\ with open-source architecture datasets like NAS-Bench-101 and NAS-Bench-301. The experimental results demonstrate that \ourtit\ can successfully find the optimal mask for a victim model within a given FLOPs constraint, leading up to 2.6\% inference accuracy degradation for attackers with only $0.14\times$ FLOPs overhead.
  The code is available at: \url{https://github.com/Tongzhou0101/ObfuNAS}.

\end{abstract}

\keywords{Deep neural network, Security, Side channels, Architecture obfuscation}

\maketitle

\section{Introduction}

The architecture of a deep neural network (DNN) plays an essential role in its performance, such as inference accuracy and latency. As a result, searching for the optimal DNN architecture has become a critical step, which is extremely costly due to the exponentially large architecture space, e.g., $10^{18}$ candidates in DARTS \cite{liu2018darts}. Therefore, high-performance neural architectures are valuable assets for model developers, and becoming the prime targets of adversarial attacks. For a concrete example, an attacker can extract the architecture of a DNN model and then train a competitive substitute model with high performance for commercial interest \cite{yu2020deepem}. Importantly, side-channel-based DNN architecture extraction, among the existing attacks, has been successfully demonstrated in various hardware platforms like GPU \cite{zhu2021hermes}, FPGA \cite{yu2020deepem}, CPU \cite{yan2020cache}, and embedded processors \cite{batina2019csi}.

Since it is difficult to fully eliminate the association between a DNN architecture and its side channels on hardware devices, a solution to mitigate the side-channel-based DNN architecture extraction attacks is to obfuscate the DNN architecture, such as the topology, layer types, and layer dimensions \cite{li2021neurobfuscator,luo2022nnrearch}. For example, NeurObfuscator \cite{li2021neurobfuscator} proposed by Li \textit{et al.} employs eight obfuscation strategies to hide the original DNN model architecture, i.e., to make it more different. 
Although these strategies can prevent accurate DNN architecture extraction by introducing prediction errors in architectural parameters like the number of layers and dimensions, they all neglect that the architecture difference should not be the only key metric to measure the obfuscation effects. 
In fact, a mask model with a large architectural difference
from the victim model can still have high, or even higher, 
inference accuracy 
and hence be of great value to an adversary. 

\begin{figure}[t]
    \centering
    \includegraphics[scale=0.4]{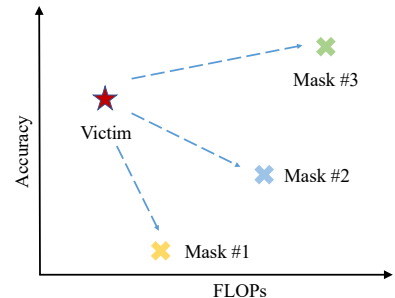}
    \caption{An illustration of different obfuscation schemes for the victim model. }
    \label{fig:illustration}
\end{figure}

\begin{figure}[htbp]

    \centering
    \includegraphics[scale=0.35]{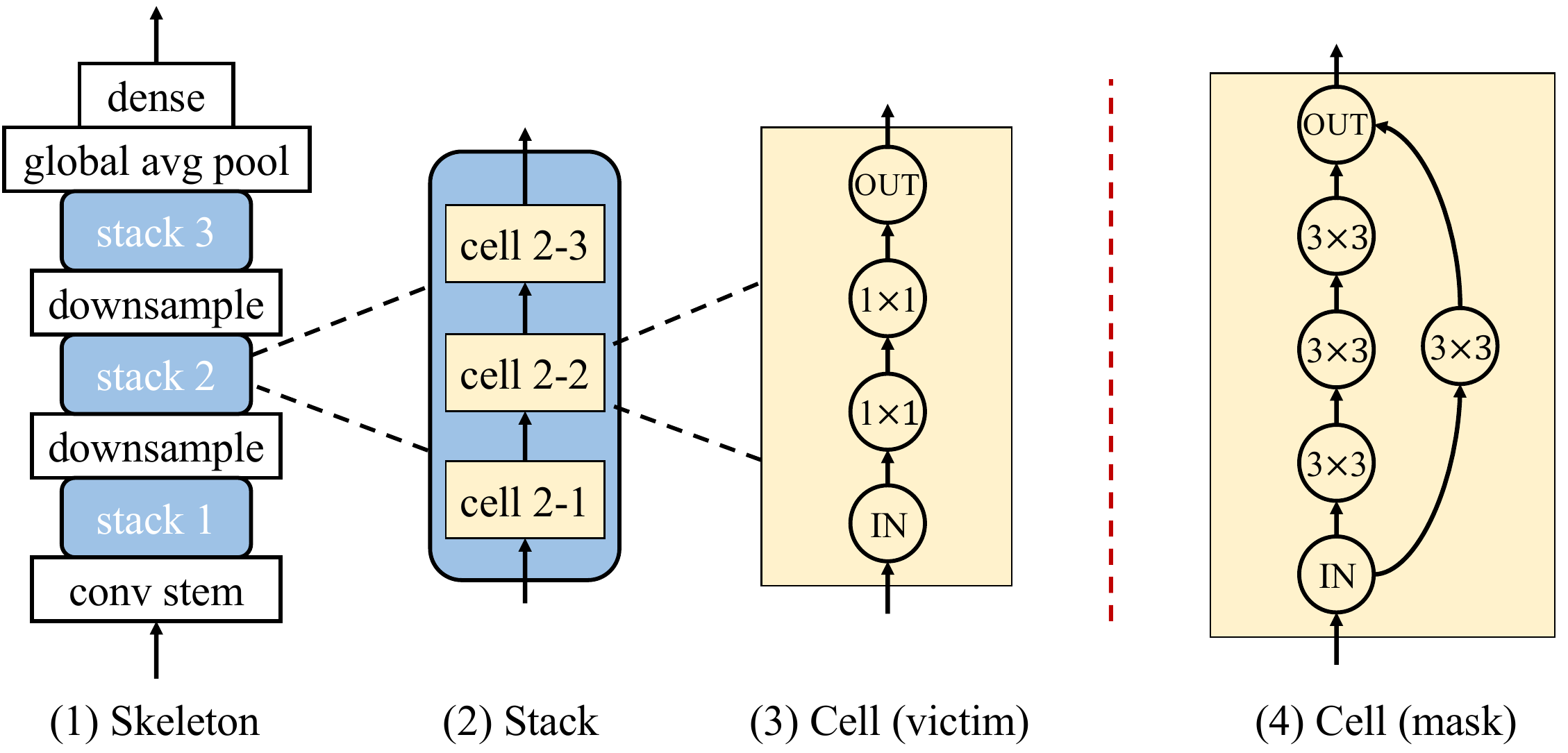}
    \caption{An example in NAS-Bench-101 \cite{ying2019bench}.}
    \label{fig:motivation}
\end{figure}

We illustrate the drawbacks of the architecture-difference oriented obfuscation schemes in Fig.~\ref{fig:illustration}. Assuming similar obfuscation strategies are applied to the victim model with different latency budgets (measured by floating-point operations, FLOPs), Mask \#3 will be selected as the optimal mask, since it allows more obfuscation space for architecture difference. If so, the victim model will be mapped accordingly to preserve its original inference accuracy, while the adversary will be capable to train the mask to reach higher inference accuracy. However, as shown in Fig.~\ref{fig:illustration}, the overall optimal mask should be Mask \#1 if the FLOPs-accuracy trade-off is considered. To further support this point, we present an example in Fig. \ref{fig:motivation}. We select a victim model in NAS-Bench-101 \cite{ying2019bench} with 77.2\% inference accuracy on CIFAR-10 \cite{cifar10_data}. If adopting the architecture-difference objective, we can get a mask with a different cell structure shown in Fig. \ref{fig:motivation} (4). However, the mask can achieve 93.02\% inference accuracy, which allows attackers to get a model with even much higher accuracy than the victim, making the obfuscation ineffective at all. 

In this work, we make the first attempt to jointly use accuracy and FLOPs as the combined obfuscation evaluation metric, for which we need to solve the following two challenges. First, like the victim architecture search space, the mask architecture search space for obfuscation is also large. Thus, manually designing mask architectures is simply out of the question. Second, apart from finding a mask architecture with low inference accuracy for obfuscation, we also need
to ensure that the adopted mask architecture does not have too large FLOPs, otherwise the victim's inference latency and energy consumption would also increase significantly. To overcome these challenges, we propose \ourtit, a neural architecture search (NAS) based DNN obfuscation approach, which aims to maximize the accuracy degradation of masks subject to FLOPs constraints. More specifically, by converting the mask architecture design into a novel FLOPs-constrained neural architecture search problem, we can leverage a well-trained super-net along with an accuracy predictor to efficiently find a mask architecture, which achieves effective obfuscation by leading the adversary to lower accuracy while meeting the FLOPs constraints.

The contributions of this work are summarized as follows:
\begin{enumerate}
   \item To the best of our knowledge, this is the first work using NAS to protect DNN against architecture extraction attacks. Leveraging the combined accuracy-FLOPs metric to guide DNN architecture masking, \ourtit~achieves more effective obfuscation than the state of the art. 
    \item We propose 7 obfuscation strategies of 3 types, which preserve the inference accuracy of the victim model. By increasing the mask architecture's training difficulties, these strategies can effectively prevent attackers from training a model with equivalent or even better performance. 
    
    \item Unlike previous obfuscation methods involving low-level modification, our proposed framework achieves DNN architecture obfuscation by only making algorithm-level changes to the victim model, which provides general applicability for any execution environment.  
\end{enumerate}

\section{Background and Related Works}

\subsection{DNN Architecture Extraction}
The performance of DNNs is largely determined by their architectures, like VGG \cite{simonyan2014very}, ResNet \cite{he2016deep}, and inception network \cite{szegedy2017inception}. Therefore, a well-designed model architecture can be considered as intellectual property with great commercial values, which motivates the architecture extraction attack. 
For example, it is demonstrated that an adversary is able to extract the DNN architecture in Machine-Learning-as-a-Service (MLaaS) platforms using the cache side channel \cite{yan2020cache}. Similarly, other side channels can also be used for such architecture extraction.  
In \cite{hua2018reverse}, Hua \etal\ successfully inferred the underlying network architectures using the memory and timing side channels during the DNN execution. Besides, Batina. \etal\ \cite{batina2019csi} utilized electromagnetic (EM) side channel to reverse-engineer the important parameters of the architecture, e.g., the number of layers and layer dimensions, to infer the victim DNN. Upon extracting the DNN architecture, attackers can further train a substitute model with competitive inference accuracy.

\subsection{DNN Architecture Protection}
Targeting these side-channel-based DNN architecture attacks, previous works have explored countermeasures from hardware platform design\cite{wang2019npufort} to DNN execution \cite{li2021neurobfuscator}. For example, Liu \etal\ proposed a method to defend architecture extraction utilizing memory access patterns, which involves oblivious shuffle, address space layout randomization, and dummy memory accesses \cite{liu2019mitigating}. Luo \etal\ proposed a framework to increase the difficulty of extracting DNN architectures from EM side-channel leakage through scheduling the tensor program execution \cite{luo2022nnrearch}. Besides, NeurObfuscator is proposed to prevent exact architecture extraction \cite{li2021neurobfuscator} through obfuscating the original dimension and number of the victim DNN layers. 
However, these methods require low-level modification and are limited by the execution environment. More importantly, existing works failed to take into account the inference accuracy of the mask architecture --- by extracting the obfuscated mask architecture, the attacker can still obtain high, or even higher, inference accuracy than the victim. 
\begin{figure*}[htbp]
    \centering
    \includegraphics[scale=0.64]{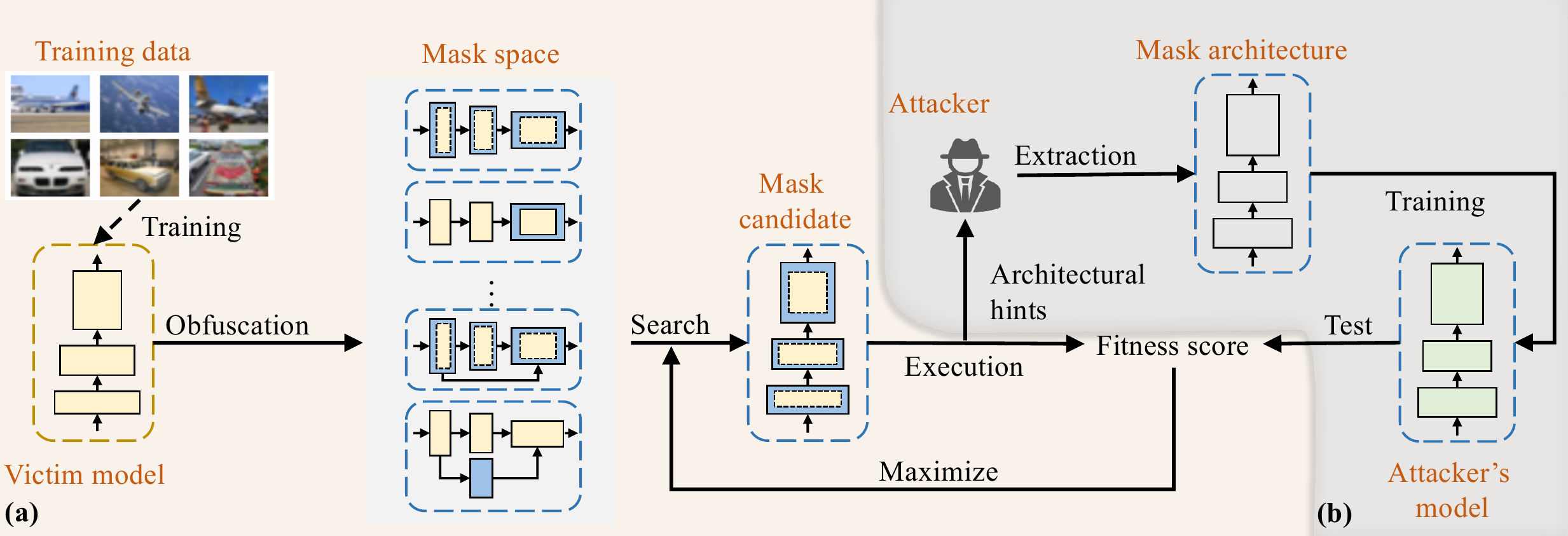}
    \caption{The overview of \ourtit. (a) Protection workflow: apply obfuscation strategies to the victim model and search the optimal mask causing the maximal accuracy degradation in mask space, and then the mask will be executed during inference.  (b) Attack workflow: an attacker will extract the model architecture by leveraging the architectural hints during victim execution, then train the extracted architecture with 
   similar training data as used in the victim model. The fitness score will be calculated later to guide the mask search.}
    \label{fig:overview}
\end{figure*}

\subsection{Neural Architecture Search (NAS)}
Designing a high-performance DNN requires not only substantial time and resources, but also domain knowledge and expertise. To ease these requirements,  NAS has been recently developed to search for Pareto-optimal DNNs, i.e., those with the highest accuracy given a FLOPs (or inference latency/energy) constraint \cite{tan2019mnasnet,liu2018progressive,wen2020neural}. Such DNNs are the focus of the architecture search and most worthy of protection. 
There are three key components of NAS: search space, search strategy, and architecture evaluation \cite{he2021automl}. 
Once a new architecture is selected from the search space based on the search strategy, its performance would be evaluated to guide the NAS process, which is time-consuming and resource-intensive.  
Therefore, many techniques have been proposed to accelerate the process of
model evaluation, such as weight sharing \cite{bender2018understanding,wang2021attentivenas}.
Additionally, training an accuracy/latency performance predictor to filter out
those unlikely optimal architectures is also commonly used, where only the top performance architectures are selected for actual evaluation  \cite{wen2020neural,wei2022npenas,wang2021alphanet}. 
Last but not least, several benchmarks have been released for quick and fair comparison for NAS algorithms, which allows obtaining the network performance by querying the pre-computed dataset \cite{ying2019bench} or a surrogate model \cite{siems2020bench}.

\subsection{Threat Model}
To explore a generic defense method, we adopt a strong threat model in this work. Specifically, we assume an attacker can perfectly extract the architecture, but not the weights, of the executed DNN (e.g., victims or masks), including the topology and activation functions, through architectural hints like side-channel leakage \cite{batina2019csi,hu2020deepsniffer,wei2020leaky,zhu2021hermes}.
Moreover, we assume a strong attacker whose DNN model training ability is as strong as the victim model developer, i.e., given an architecture $\pi$, if the developer can train it and achieve inference accuracy $ACC_{\pi}$, the attacker can achieve $ACC_{\pi}$ as well with training $\pi$ on the similar training data as the developer.

\section{Proposed Approach: \ourtit}
The overview of \ourtit\ is shown in  Fig. \ref{fig:overview}, which is proposed to solve two key problems: 1) how to obfuscate the DNN architecture with the original model accuracy preserved; 2) how to achieve the best obfuscation performance, i.e., the maximal accuracy drop for attackers while satisfying a FLOPs constraint. 

For the first problem, we propose 7 obfuscation strategies of 3 categories, namely, scaling-up (Sec. \ref{sec:scaling-up}), operation-change (Sec. \ref{sec:operation-change}), and connection-adding (Sec. \ref{sec:connection-adding}). 
The principle behind these strategies is to increase the training difficulties of the mask architecture, 
leading attackers to lower inference accuracy after extracting and training the mask.
Besides, we will prove that these strategies are function-preserving to make sure that the victim model can preserve its original inference accuracy after obfuscation, as described in Eq. (\ref{eq:preserving}):
\begin{equation}\label{eq:preserving}
    \forall x: f(x|\theta_f) = g(x|\theta_g),
\end{equation}
where $x$ denotes the input of the network, $f$ represents the victim network, $g$ represents the obfuscated network, and  $\theta$ is the corresponding network parameters. With proposed strategies, we can generate a mask space for the victim model.

As for the second problem, we will simulate the attacking process and use its result to guide the optimal mask search (see Sec. \ref{sec:search}). Considering obfuscation is at the cost of the FLOPs budget, we propose resource-constrained search, i.e., search for the optimal mask within a given FLOPs constraint, which is implemented by an evolutionary search with accuracy drop as the fitness score.

\subsection{Scaling-up Obfuscation}
\label{sec:scaling-up}
It is well-known that the optimization/training difficulties will grow with the increase of DNN dimensions \cite{srivastava2015training}. 
Based on this observation, we propose 3 obfuscation strategies by scaling up the victim DNN architecture to increase training difficulties for the attackers from width, depth,  and kernel size, respectively.

\textbf{Layer Widening.}
Layer widening is proposed to increase the output dimension of a layer. Since it is commonly applied to the convolutional (Conv) layer, here we will use Conv to illustrate this strategy. Suppose the weights of a Conv layer $L$ is $W^{(L)}\in \mathbb{R}^{k_1,k_2,i,o}$, where $k_1\times k_2$ is the kernel size (we assume the kernel sizes of all Conv layers are the same for simplicity), $i$ denotes the number of input channels, and $o$ stands for the number of output channels. After layer widening, the weights of $L$ will be  $V^{(L)}\in \mathbb{R}^{k_1,k_2,i,o'}$, with $o'>o$, and the weights of the subsequent Conv layer $L+1$ would change from $W^{(L+1)}\in \mathbb{R}^{k_1,k_2,o,q}$ to $V^{(L+1)}\in \mathbb{R}^{k_1,k_2,o',q}$ to match the increased output channels of $L$. 

To preserve the function of the original layers, we need to adjust the value of  $V^{(L)}$ and $V^{(L+1)}$ to satisfy the following equations:

\begin{equation}
    V^{(L)}_{\cdot ,\cdot ,\cdot ,j} = \left\{\begin{matrix}
 W^{(L)}_{\cdot ,\cdot ,\cdot ,j},& j\leq o  \\
 0,&  o<j\leq o'\\
\end{matrix}\right. ,
\end{equation}
\begin{equation}
    V^{(L+1)}_{\cdot ,\cdot ,t,\cdot} = \left\{\begin{matrix}
 W^{(L+1)}_{\cdot ,\cdot ,t,\cdot},& t\leq o  \\
 random,&  o<t\leq o'
\end{matrix}\right. .
\end{equation}

Note that this strategy can also work for the fully connected layer, which can be replaced with a Conv layer with kernel size 1$\times$1.

\textbf{Layer Deepening.} We use layer deepening to increase the depth of DNNs by sequentially inserting additional layers. To be function-preserving, the inserted layer should function as an identity layer, i.e., the input of this layer is equal to its output. If the inserted layer is a fully connected layer, we can simply set its weights to an identity matrix. Otherwise, suppose we insert a Conv layer with weights $U\in \mathbb{R}^{k_1,k_2,o,o}$ between two sequential Conv layers $W^{(L)}\in \mathbb{R}^{k_1,k_2,i,o}$  and $W^{(L+1)}\in \mathbb{R}^{k_1,k_2,o,p}$, then $U$ should be set to: 
\begin{equation}
U_{p,q,t,j}=\left\{\begin{matrix}
1, & p=\frac{k_1+1}{2} \wedge q=\frac{k_2+1}{2}\wedge t=j\\
0, & otherwise \\
\end{matrix}\right. .
\label{eq:identity}
\end{equation}

Besides, if the inserted layer is followed by an activation function $\phi(\cdot)$, it should satisfy the restriction $\phi(\cdot) = \phi(\phi(\cdot))$, e.g., the commonly used activation function ReLU. Moreover, extra efforts are required if batch normalization is used. Specifically, batch normalization will do the following transformation during inference \cite{ioffe2015batch}:
\begin{equation}
y=\frac{\gamma }{\sqrt{Var[x]+\epsilon }}\cdot x+(\beta -\frac{\gamma E(x) }{\sqrt{Var[x]+\epsilon }}),
\end{equation}
where $\epsilon$ is a small self-define value, $E(x)$ and $Var[x]$ are the mean and variance of input data, which are fixed during inference. Therefore, by setting $\gamma= \sqrt{Var[x]+\epsilon }$ and $\beta=E(x)$, we can undo the normalization and successfully build an identity layer.

\textbf{Kernel Widening.} 
A Conv layer with weights $W\in \mathbb{R}^{k_1,k_2,i,o}$ after kernel widening would become $U\in \mathbb{R}^{k_3,k_4,i,o}$, where $k_3>k_1$ and $k_4>k_2$. The function preservation is straightforward, i.e.,
\begin{equation}
U_{p,q,\cdot,\cdot}=\left\{\begin{matrix}
W_{p,q,\cdot,\cdot}, & \frac{k_3-k_1}{2}\leq p \leq \frac{k_3+k_1}{2} \wedge \frac{k_4-k_2}{2}\leq q \leq \frac{k_4+k_2}{2}\\
0, & otherwise \\
\end{matrix}\right. .
\end{equation}
Besides, zero-padding can be applied to the input feature maps in order to preserve the original size of output feature maps after convolution.

\subsection{Operation-change Obfuscation}
\label{sec:operation-change}
Since Conv layers are functional, we aim to replace some non-parameter layers with it to increase the number of trainable parameters, leading to an increase in training difficulties of the obfuscated architecture. Such replacement is described as the operation-changing obfuscation strategy in this work. 
The principle of the non-parameter layer selection is that the layer can be represented by a Conv layer with function preserved. Following this principle, we select two common DNN operations, average pooling and skip connection.

\textbf{Average Pooling Replacement.} Suppose a Conv layer uses the same kernel size (${k_1\times k_2}$), stride, and padding pattern as the average pooling layer, then it can work as average pooling when setting every value of its weights to $\frac{1}{k_1\times k_2}$.

\begin{figure}[htbp]

\centering
\includegraphics[scale=0.5]{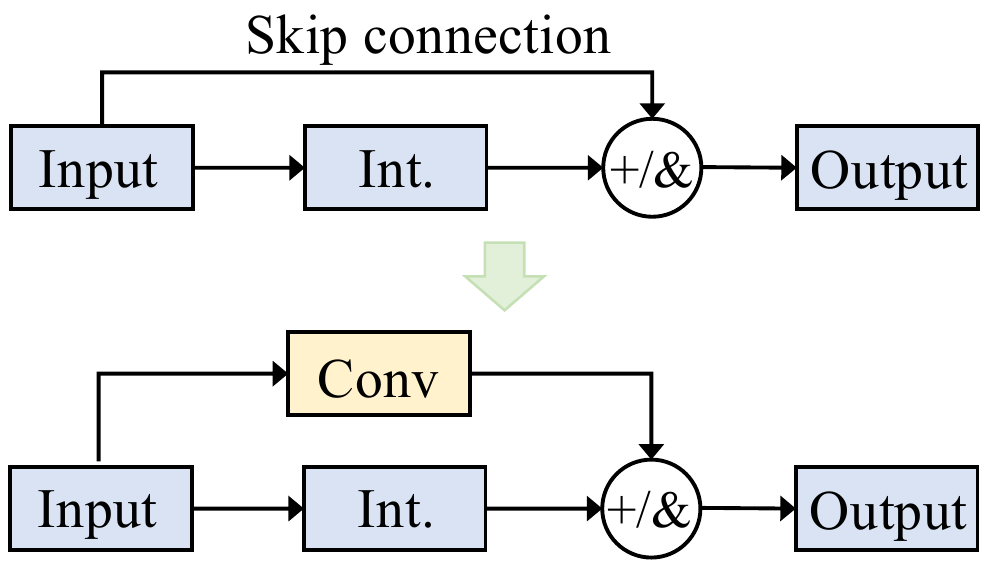}
\caption{Skip connection replacement.}
\label{fig:skip}
\end{figure}

\textbf{Skip Connection Replacement.} Since skip connection is functionally equal to an identity layer, this replacement can be transferred to using a Conv layer to perform identity operation, as shown in Fig. \ref{fig:skip}, where Int. refers to the intermediate layer(s), and +/\& indicates sum/concatenate operation. Thus, the weights adjustment of this Conv layer is the same as Eq. (\ref{eq:identity}).

\subsection{Connection-adding Obfuscation}\label{sec:connection-adding}
In this category, the original connection will be preserved to maintain the inference accuracy. Moreover, we add extra connections to disturb the signal propagation, which will also increase the optimization difficulties for the attacker while training the extracted mask. 

\begin{figure}[htbp]
\centering
 \includegraphics[scale=0.56]{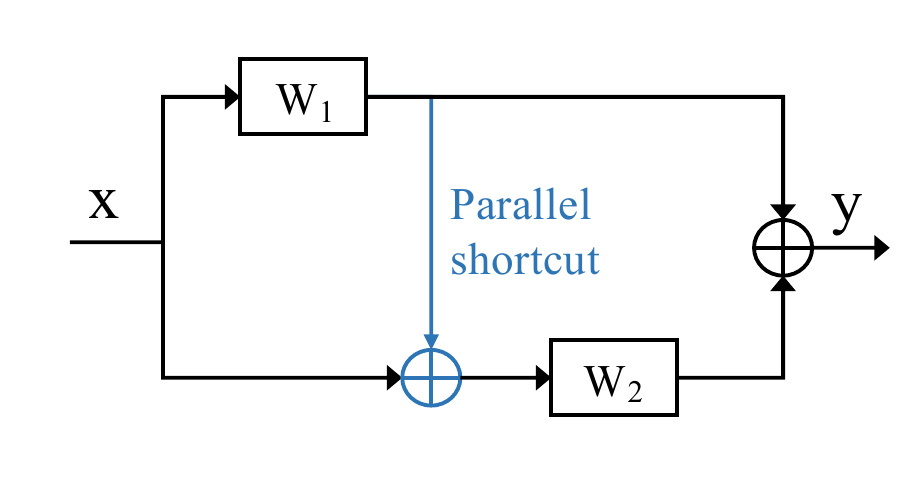}
    \caption{Parallel shortcut adding.}
    \label{fig:parallel}
\end{figure}

\textbf{Shortcut Adding.} This strategy is used to add a shortcut between two non-directly connected layers if their dimensions match. The added shortcut can be further divided into the sequential shortcut and the parallel shortcut, depending on the hierarchy of these two layers. Specifically, if these two layers are sequential, the added shortcut will be the same as the skip connection in Fig. \ref{fig:skip}. 

As for the parallel shortcut, 
one example is shown in Fig. \ref{fig:parallel}, where $x$ is the input, $y$ is the output, $W_1$ and $W_2$ are the weights of two Conv layers. 
Before adding the parallel shortcut, we have 
\begin{equation}
\label{eq:no_para}
\begin{split}
    y = W^T_1 * x &+ W^T_2 * x,\\
    \frac{\partial y}{\partial W_1} = & \frac{\partial y}{\partial W_2} = x,
\end{split}
\end{equation}
while after adding the parallel shortcut, Eq. (\ref{eq:no_para}) will become
\begin{equation}
\label{eq:para}
\begin{split}
    y = W^T_1 * x &+ W^T_2 (W^T_1 * x + x),\\
    \frac{\partial y}{\partial W_1} = x + W^T_2 * x, & \quad \frac{\partial y}{\partial W_2} = x + W^T_1 * x,
\end{split}
\end{equation}
which indicates that the parallel shortcut adding will cause the gradient update of $W_1$ and $W_2$ influenced by each other. As a result, the optimization difficulties of architectures with such connections would increase. To preserve the functionality, the shortcut feature maps should be multiplied by 0 before summation.
                                              
\textbf{Layer Branch Adding.} The connection of the layer branch adding is similar to the shortcut adding strategy, except that the shortcut will be replaced by an operation to increase the model complexity. Besides, the chosen operation should also preserve the size of feature maps, so the feature maps can be later added together. For this strategy, the function preservation can be achieved by setting the weights of the operation to 0.

\begin{table*}[htbp]
\centering

\caption{The search space of AlphaNet.}
\label{tab:alphanet_space}
\renewcommand\tabcolsep{22pt}

\begin{tabular}{ccccc}
\toprule
Block  &   Width            & Depth           & Kernel size & \quad Expansion ratio \\ \hline
Conv     & \{16,24\}           & -               & 3           & -               \\
MBConv-1 & \{16,24\}           & \{1,2\}         & \{3,5\}     & 1               \\
MBConv-2 & \{24,32\}           & \{3,4,5\}       & \{3,5\}     & \{4,5,6\}       \\
MBConv-3 & \{32,40\}           & \{3,4,5,6\}     & \{3,5\}     & \{4,5,6\}       \\
MBConv-4 & \{64,72\}           & \{3,4,5,6\}     & \{3,5\}     & \{4,5,6\}       \\
MBConv-5 & \{112,128\}         & \{3,4,5,6,7,8\} & \{3,5\}     & \{4,5,6\}       \\
MBConv-6 & \{192,200,208,216\}  & \{3,4,5,6,7,8\} & \{3,5\}     & \{4,5,6\}       \\
MBConv-7 & \{216,224\}         & \{1,2\}         & \{3,5\}     & 6               \\
MBPool   & \{1792,1984\}       & -               & 1           & 6               \\ \bottomrule
\end{tabular}
\end{table*}

\subsection{Optimal Mask Search}
\label{sec:search}
In this section, we propose resource-constrained search and apply the evolutionary algorithm to efficiently search for the optimal mask.

\textbf{Resource-constrained Search.} Our objective is to search for a mask that yields the maximal accuracy drop with FLOPs constrained, which can be formulated as follows: 
\begin{equation}
\label{eq:objective_2}
\begin{split}
  \pi^{\ast } =\mathop{\arg\max}_{\pi_i \in \mathbf{\Omega}} \mathcal{L} & (W_{\pi_i}; X_{val}),    \\
  s.t.\ FLOPs &(\pi_i)  <\tau,
\end{split}
\end{equation}
where $W_{\pi_i}$ is the network parameters associated with the mask $\pi_i$,  $\mathbf{\Omega}$ denotes the mask space, $X_{val}$ denotes the validation dataset, $\mathcal{L}(\cdot)$ stands for the loss function, $\pi^{\ast }$ is the optimal mask we expect to find out, and $\tau$ is the given FLOPs constraint.

\textbf{Evolutionary Algorithm.}  
In this work, we adopt an evolutionary algorithm that has been demonstrated to be effective for NAS problems \cite{liu2017hierarchical,real2019regularized,suganuma2017genetic}. Specifically, 
our fitness function F is defined as:
\begin{equation}
    F(\pi_i) = -ACC_{val}(\pi_i),
    \label{eq:fitness}
\end{equation}
where $ACC_{val}(\cdot)$ is the validation accuracy of a mask. The searching process aims to find the mask with the highest fitness score within a certain FLOPs constraint, which is consistent with Eq. (\ref{eq:objective_2}).

\section{Experimental Validation}
We evaluate the performance of \ourtit\ based on three architecture spaces used in AlphaNet \cite{wang2021alphanet}, NAS-Bench-101 \cite{ying2019bench}, and NAS-Bench-301 \cite{siems2020bench} (Sec \ref{sec:space}). Specifically, in each space, we select several Pareto-optimal architectures as the victim models and adopt applicable obfuscation strategies (Sec. \ref{sec:setup}). We then search for the best mask with the maximal accuracy degradation for each victim model and compare the results with the state-of-the-art (SOTA)  obfuscation framework, i.e., NeurObfuscator (Sec. \ref{sec:result}). 
Since our approach focuses on algorithm-level obfuscation, we exclude two obfuscation knobs in NeurObfuscator, i.e., optimization knobs and scheduling knobs, for fair comparisons. 

\subsection{\textbf{Architecture Search Space}}
\label{sec:space}

\subsubsection{\textbf{AlphaNet.}}
AlphaNet aims to search for sub-nets from a super-net that can achieve Pareto-optimal performance on ImageNet. The search space of the super-net is defined in Table~\ref{tab:alphanet_space}, where MBConv denotes the inverted residual block used in \cite{sandler2018mobilenetv2}, the expansion ratio is the parameter of the Conv layer inside MBConv, and MBPool is the last Conv layer with average pooling. Moreover, since AlphaNet provides a well-trained super-net, we can build an architecture dataset consisting of its sub-nets and evaluate them to obtain their inference accuracy. 

\subsubsection{\textbf{NAS-Bench-101 (NB-101).}}
NB-101 provides a public architecture dataset for NAS, including 423k unique DNN architectures 
and 5M models trained and evaluated on CIFAR-10. All unique architectures are made up of the same number of cells 
but with different cell structures, which is represented by a directed acyclic graph (DAG) with up to 7 nodes and 9 edges. In a DAG,
each node and edge indicate an operation and a feature tensor, respectively. Apart from the input and output nodes, others have 3 possibilities: $3\times3$ convolution, $1\times1$ convolution, and $3\times3$ max pooling.

\subsubsection{\textbf{NAS-Bench-301 (NB-301).}}
NB-301 is a surrogate NAS benchmark that includes about $10^{18}$ architectures covered by the DARTS search space \cite{liu2018darts}. The DNN backbone of all architectures is built with 8 cells categorized into two types, namely normal cell and reduction cell. Each cell can be represented as a DAG with 12 edges and 7 nodes (2 input nodes, 4 intermediate nodes, and 1 output node). Here, each node and edge indicate a feature tensor and an operation, which is different from the definition in NB-101. Each intermediate node is the addition result of two operations, and all results of intermediate nodes will be concatenated as the final output. There are 7 operations involved: 3$\times$3/5$\times$5 separable/dilated convolution, 3$\times$3 average/max pooling, and identity. Besides, NB-301 includes a trained surrogate model to predict the accuracy performance on CIFAR-10 for each architecture in the search space.

\subsection{\textbf{Experimental Setup}}
\label{sec:setup}
\begin{figure}[htbp]
\centering
\includegraphics[scale=0.4]{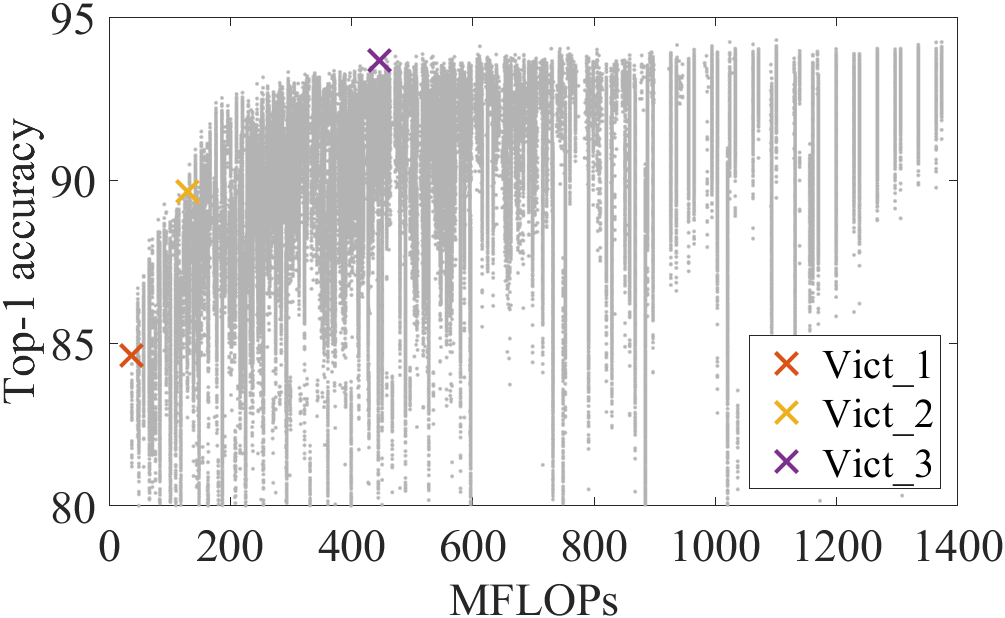}
\caption{Victims in NB-101.}
\label{fig:victim_101}
\end{figure}

\begin{figure}[htbp]
\centering
\includegraphics[scale=0.4]{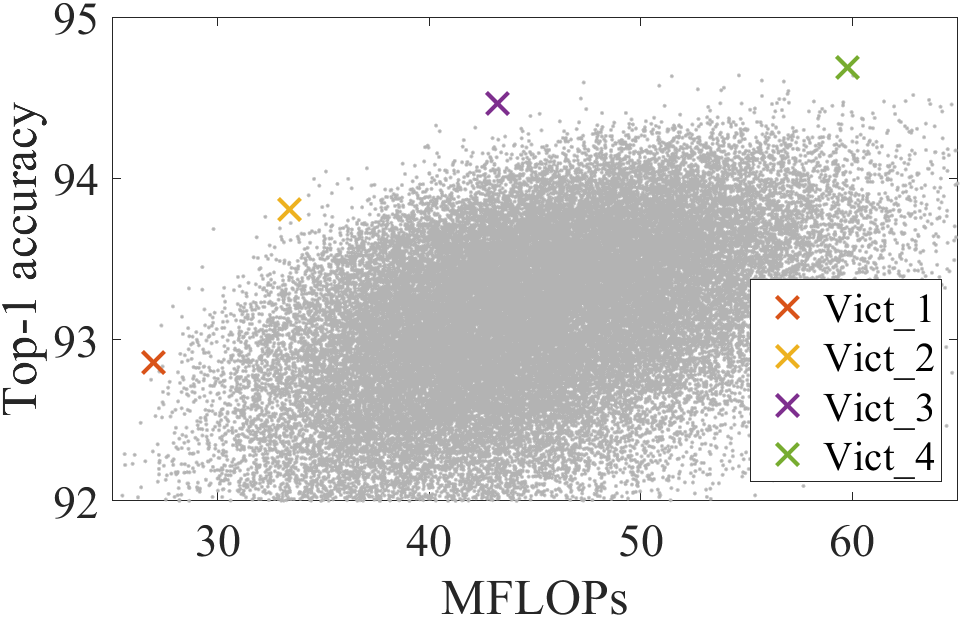}
\caption{Victims in NB-301.}
\label{fig:victim_301}
\end{figure}

Since AlphaNet has reported 8 Pareto-optimal architectures with different FLOPs \cite{wang2021alphanet}, i.e., A0-A6 (A5 has two versions) shown in Table \ref{tab:res}, we directly use them as the victim models for further obfuscation. However, for NB-101 and NB-301 that do not provide any Pareto-optimal architectures, we select a few architectures with different accuracy-FLOPs trade-offs on the Pareto front as the victim models, with selected architectures shown in Fig. \ref{fig:victim_101} and Fig. \ref{fig:victim_301}, respectively.

Due to the difference in the search space of these three datasets, the applicable obfuscation strategies also vary, see Table \ref{tab:strategies}, where the checkmark denotes all strategies in that category are applicable. Specifically, obfuscation for AlphaNet is straightforward, i.e., scaling up the dimensions from different perspectives within the defined search space. Specially, the activation function used in AlphaNet is swish \cite{wang2021alphanet}, defined as 
\begin{equation}
    swish(x)=x*\frac{1}{1+e^{-x}},
\end{equation}
where x is the input feature map.
To satisfy the restriction $\phi(\cdot) = \phi(\phi(\cdot))$ in layer deepening, we construct a fake swish function with the same operators (e.g., addition and division) to replace the original swish function for added layers, i.e.,
\begin{equation}
    t = 1+e^{-x}, \quad fake\_swish(x)=x*\frac{t}{t}.
\end{equation}
Following the conclusions in \cite{batina2019csi}, an attacker can deduce the activation function by comparing its timing side-channel leakage (i.e., time duration) with other known activation functions, the proposed self-defined fake swish will still be identified as swish due to the similar operations.

\begin{table}[htbp]
\centering
\caption{Obfuscation strategies for each architecture dataset. OP-change denotes operation-change and CN-adding denotes connection-adding obfuscation.}
\label{tab:strategies}
\renewcommand\tabcolsep{5pt}
\begin{tabular}{cccc}
\toprule
    & Scaling-up     & OP-change    & CN-adding \\ \midrule
AlphaNet     & \checkmark &   -      & -         \\ 
NB-101 & Depth, Kernel&  - &  \checkmark   \\ 
NB-301 & Kernel & \checkmark &  -   \\ \bottomrule
\end{tabular}
\end{table}

\begin{figure}[htbp]
    \centering
    \includegraphics[scale=0.38]{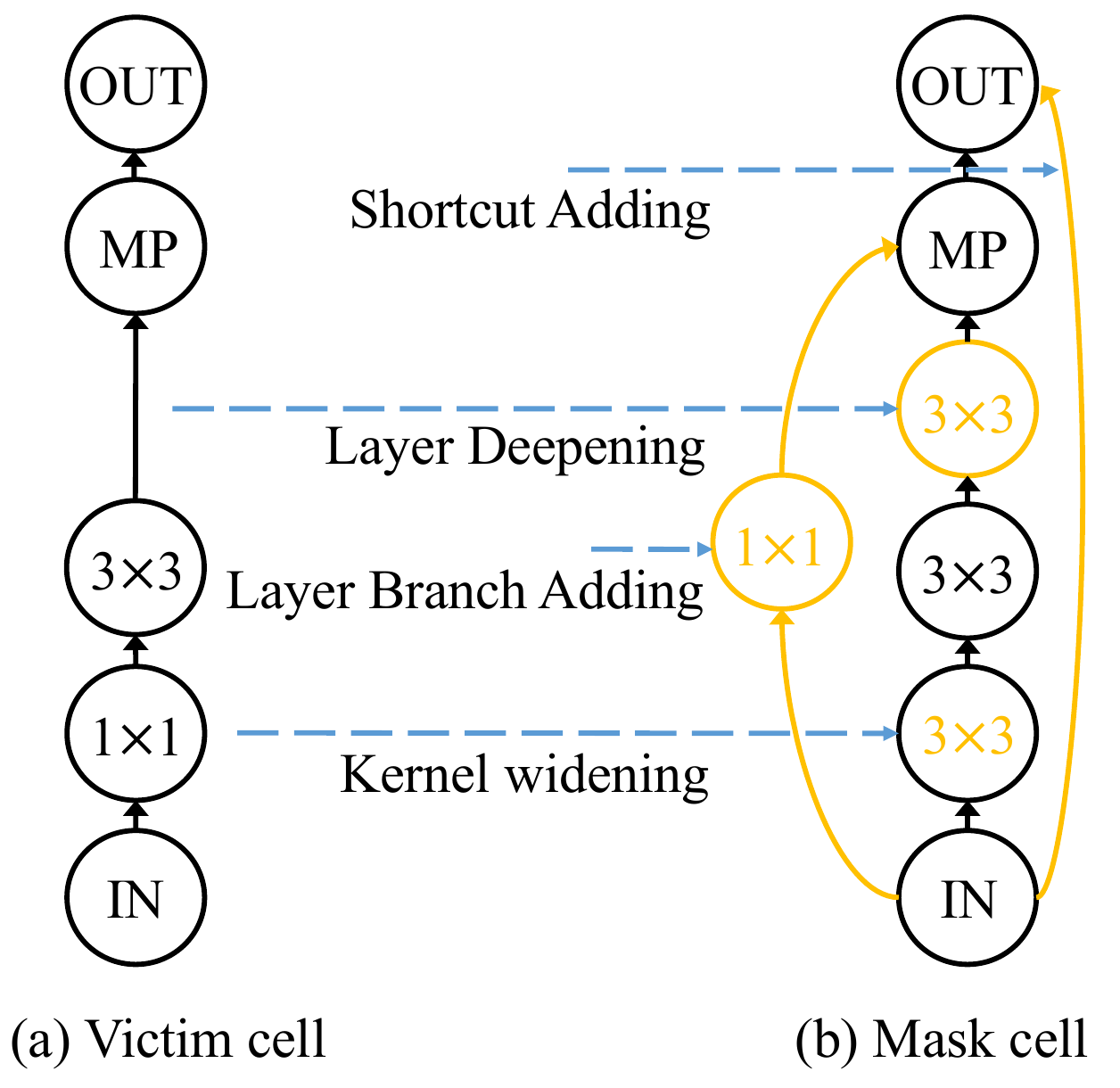}
    \caption{An Illustration of obfuscation strategies for cells in NB-101, where IN and OUT denote input and output feature maps, $1\times1$ and $3\times3$ are the kernel sizes of Conv layers, and MP denotes $3\times3$ max pooling. The feature map size inside the cell is fixed by adding zero paddings.}
    \label{fig:obfu_101}
\end{figure}
For NB-101, other than scaling up the depth and kernel size by layer-deepening and kernel-widening strategies, the connection-adding strategies are also applicable, with an illustration shown in Fig. \ref{fig:obfu_101}. As for NB-301, the victim cells can be obfuscated with the kernel widening and the operation-change strategies.

\subsection{\textbf{Results}}
\label{sec:result}
\begin{table*}[htbp]
\centering
\caption{The obfuscation results without latency constraints.}
\label{tab:res}
\renewcommand\tabcolsep{16pt} 
\begin{tabular}{cccccccc}
\toprule
\multirow{2}{*}{} &
  \multicolumn{1}{c}{\multirow{2}{*}{\quad Victim \quad}} &
  \multicolumn{2}{c}{Original} &
  \multicolumn{2}{c}\ourtit\ &
  \multicolumn{2}{c}{NeurObfuscator \cite{li2021neurobfuscator}} \\ \cmidrule(r){3-4} \cmidrule(r){5-6} \cmidrule(r){7-8}
 & 
  \multicolumn{1}{c}{} &
  Acc &
  \multicolumn{1}{c}{MFLOPs} &
  Acc &
  \multicolumn{1}{c}{MFLOPs} &
  Acc &
  MFLOPs \\ \midrule
\multirow{8}{*}{\begin{tabular}[c]{@{}c@{}}AlphaNet\\ \\ (ImageNet)\end{tabular}} & 
  A0 &
  77.78 &
  203.39 &
  \textbf{77.41} &
  \textbf{266.50} &
  77.96 &
  342.43 \\
 &
  A1 &
  78.90 &
  279.24 &
  78.91 &
  388.01 &
  79.38 &
  498.97 \\
 &
  A2 &
  79.09 &
  316.73 &
  79.25 &
  414.36 &
  79.72 &
  479.91 \\
 &
  A3 &
  79.46 &
  356.52 &
  \textbf{79.43} &
  \textbf{433.49} &
  79.90 &
  553.56 \\
 &
  A4 &
  80.01 &
  443.53 &
  \textbf{80.00} &
  \textbf{581.05} &
  80.41 &
  651.96 \\
 &
  A5 &
  80.27 &
  491.48 &
  \textbf{80.14} &
  \textbf{606.61} &
  80.68 &
  743.34 \\
 &
  A5\_1 &
  80.74 &
  594.79 &
  \textbf{80.66} &
  \textbf{680.74} &
  81.03 &
  837.46 \\
 &
  A6 &
  80.82 &
  709.01 &
  80.85 &
  795.06 &
  81.23 &
  983.28 \\ \midrule
\multirow{3}{*}{\begin{tabular}[c]{@{}c@{}}NB-101\\  (CIFAR-10)\end{tabular}} &
  Vict\_1 &
  84.61 &
  37.70 &
  \textbf{82.65} &
  \textbf{37.70} &
  87.11 &
  202.51\\
 &
  Vict\_2 &
  89.63 &
  128.85 &
  \textbf{87.02} &
  \textbf{146.29} &
  91.73 &
  548.96 \\
 &
  Vict\_3 &
  93.67 &
  446.62 &
  \textbf{92.94} &
  \textbf{446.62} &
  93.53 &
  531.89 \\ \midrule
\multirow{4}{*}{\begin{tabular}[c]{@{}c@{}}NB-301\\  (CIFAR-10)\end{tabular}} &
  Vict\_1 &
  92.86 &
  26.95 &
  \textbf{91.74} &
  \textbf{40.46} &
  92.87 &
  27.86 \\
 &
  Vict\_2 &
  93.81 &
  33.40 &
  \textbf{92.46} &
  \textbf{49.23} &
  93.76 &
  35.43 \\
 &
  Vict\_3 &
  94.46 &
  43.24 &
  \textbf{93.14} &
  \textbf{56.64} &
  94.41 &
  44.16 \\
 &
  Vict\_4 &
  94.69 &
  59.76 &
  \textbf{93.68} &
  \textbf{71.84} &
  95.65 &
  63.43 \\ \bottomrule
\end{tabular}
\end{table*}

\begin{figure*}[htbp]
     \centering
          \begin{subfigure}{0.30\textwidth}
         \centering
         \includegraphics[width=\textwidth]{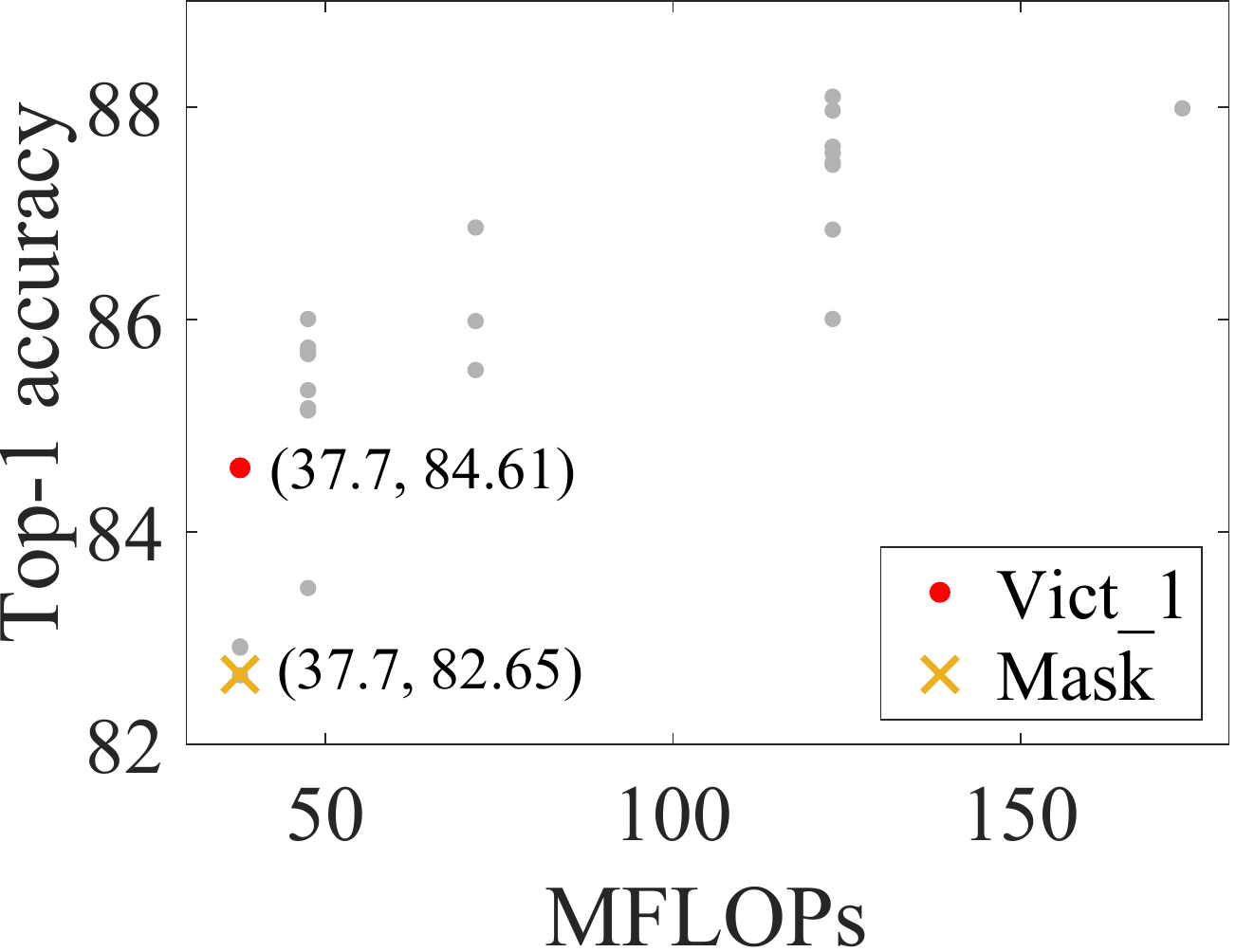}
         \caption{Masks of Vict$\_$1.}
         \label{fig:victim101_mask1}
     \end{subfigure} 
     \hspace{2em}
     \begin{subfigure}{0.31\textwidth}
         \centering
         \includegraphics[width=\textwidth]{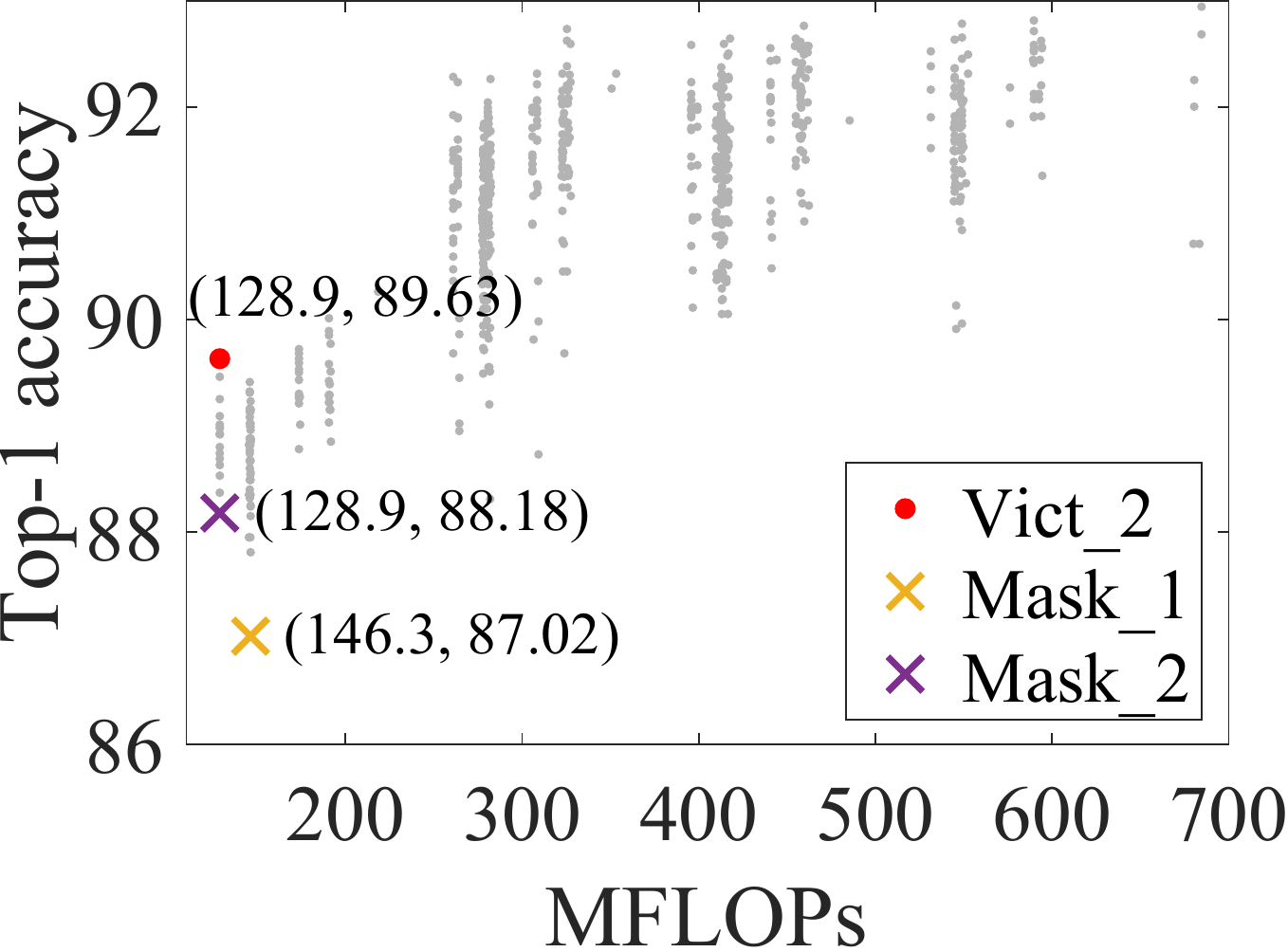}
         \caption{Masks of Vict$\_$2.}
         \label{fig:victim101_mask2}
     \end{subfigure}
    \hspace{1.5em}
     \begin{subfigure}{0.30\textwidth}
         \centering
         \includegraphics[width=\textwidth]{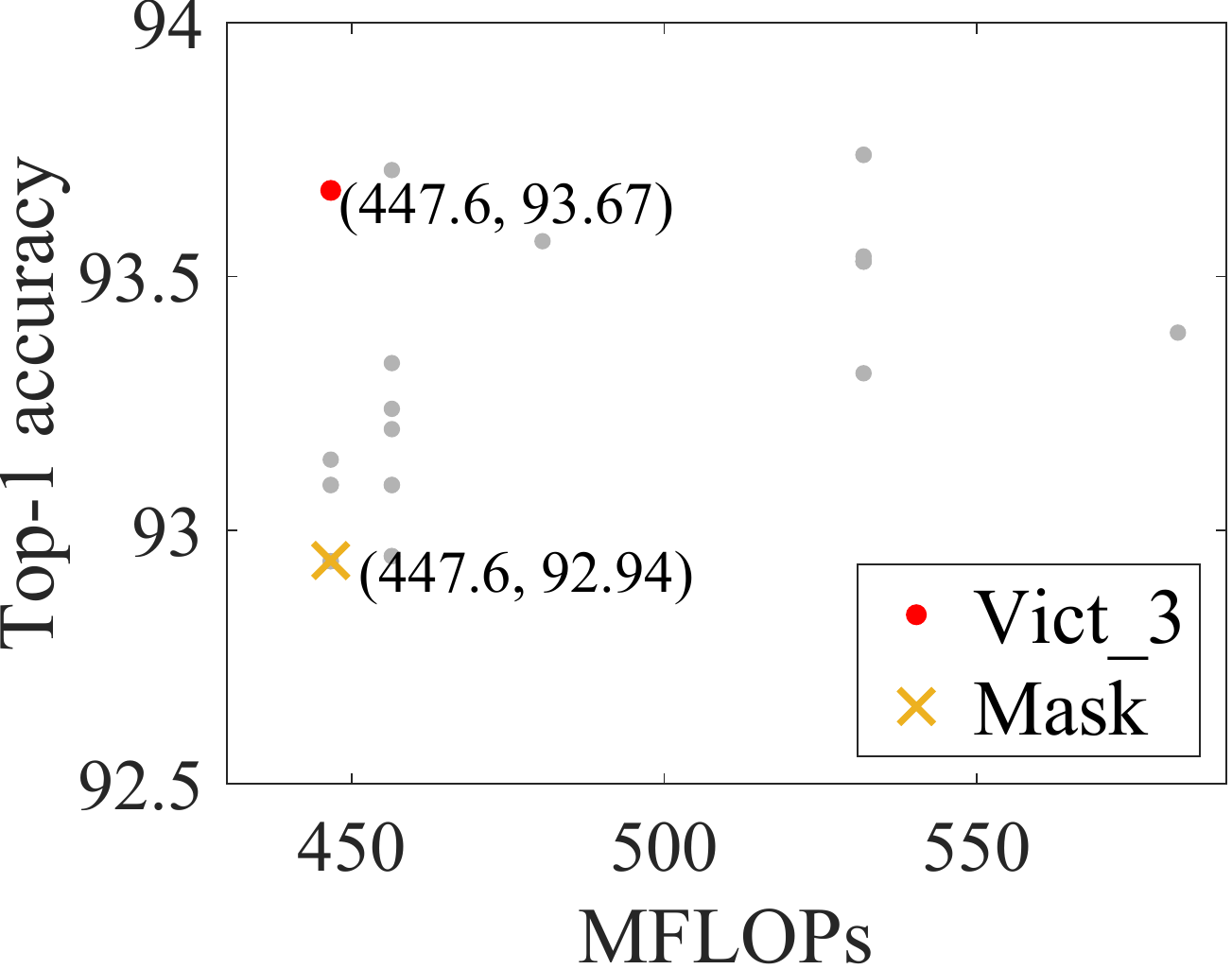}
         \caption{Masks of Vict$\_$3.}
         \label{fig:victim101_mask3}
     \end{subfigure}
        \caption{The optimal mask with different FLOPs overhead for victims in NB-101. }
        \label{fig:victim101_mask}
\end{figure*}

To explore the best obfuscation performance of \ourtit, we first relax the constraint 
to find the best masks for all victim models. As shown in Table \ref{tab:res}, the obfuscation performance of AlphaNet is not as good as expected, i.e., the accuracy of the mask is similar to the victim itself. One explanation is that the increased optimization difficulties raised by network scaling-up are limited, especially for DNNs like AlphaNet including the residual connection, which can ease the training difficulties for larger DNNs. However, our method still outperforms the SOTA (NeurObfuscator \cite{li2021neurobfuscator}), which results in higher accuracy growth with larger FLOPs overhead.

For victims in NB-101, \ourtit\ achieves around $1\%\sim3\%$ accuracy degradation with a small FLOPs overhead ($<0.14\times$), thanks to the shortcut adding strategy introducing no latency overhead. In contrast, NeurObfuscator even boosts the inference accuracy of masks up to 2.5\%. Besides, \ourtit\ protects the victims in NB-301 by deteriorating $\sim$1\% accuracy, while NeurObfuscator can only achieve at most 0.05\% accuracy drop. Since the obfuscation strategies for NB-301 involve operation-change, which replaces non-parameter operations with Conv, the FLOPs overhead, in this case, has obvious growth. Overall, \ourtit\ achieves better obfuscation performance than the SOTA obfuscation framework. Note that even $1\%$ accuracy matters a lot in NAS, thus the accuracy degradation achieved by \ourtit\ will make
the architecture extraction attacks meaningless.

\begin{figure*}[htbp]
     \centering
     \captionsetup{width=\linewidth}

     \begin{subfigure}{0.4\textwidth}
         \centering
         \includegraphics[width=\textwidth]{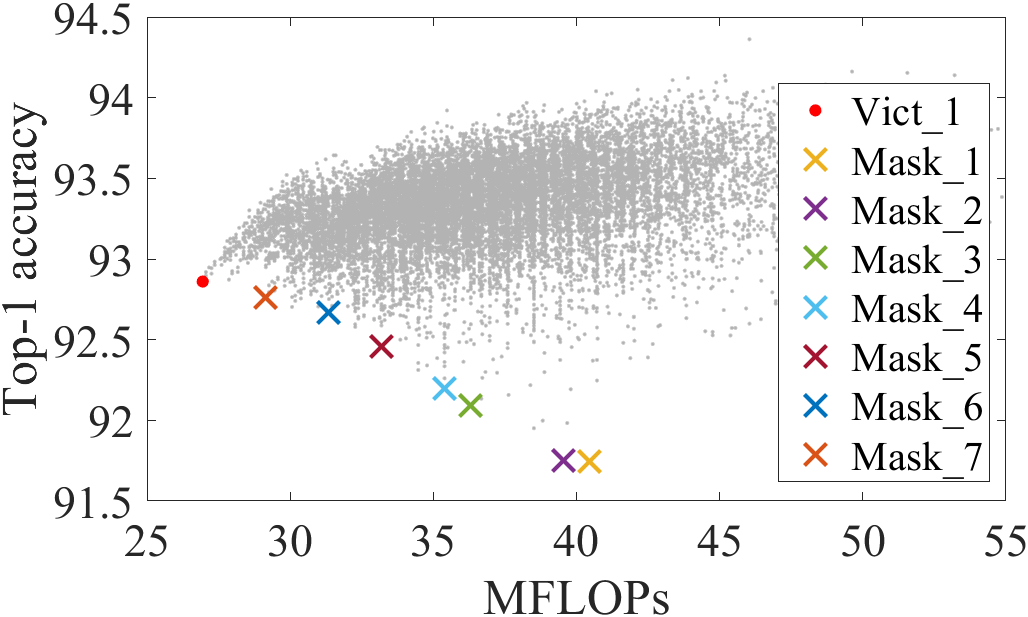}
         \caption{Masks of Vict$\_$1.}
         \label{fig:victim301_mask1}
     \end{subfigure}
  \hspace{5em}
     \begin{subfigure}{0.4\textwidth}
         \centering
         \includegraphics[width=\textwidth]{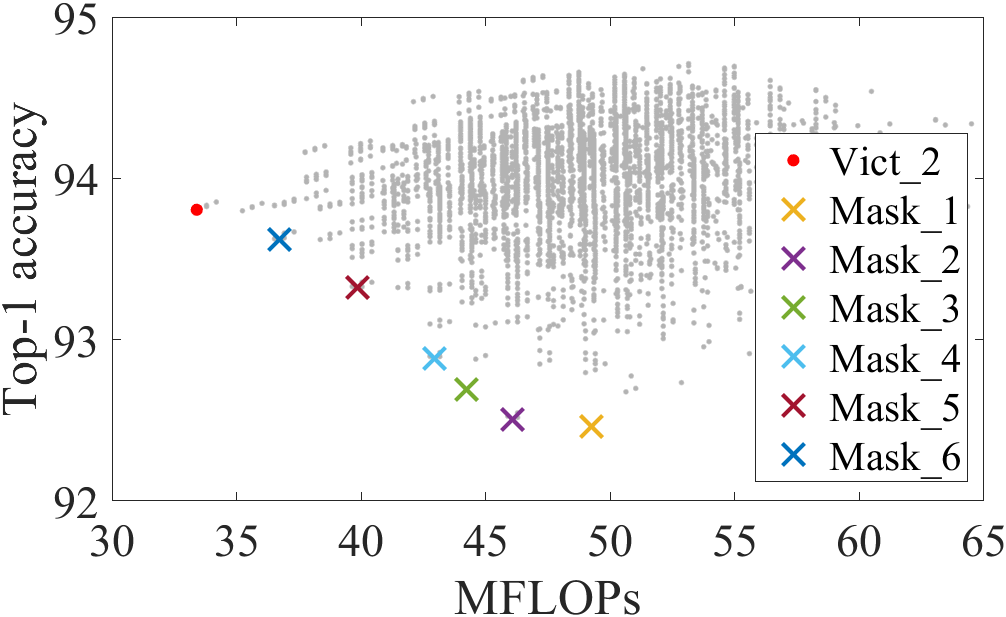}
         \caption{Masks of Vict$\_$2.}
         \label{fig:victim301_mask2}
     \end{subfigure}
   \quad
     \begin{subfigure}{0.4\textwidth}
         \centering
         \includegraphics[width=\textwidth]{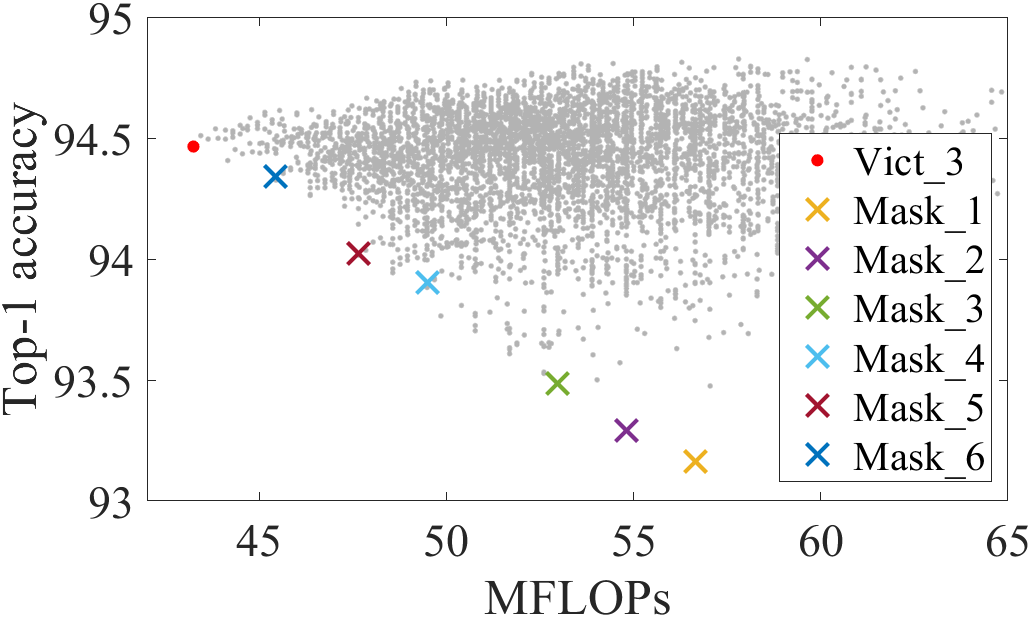}
         \caption{Masks of Vict$\_$3.}
         \label{fig:victim301_mask3}
     \end{subfigure}
  \hspace{5em}
     \begin{subfigure}{0.41\textwidth}
         \centering
         \includegraphics[width=\textwidth]{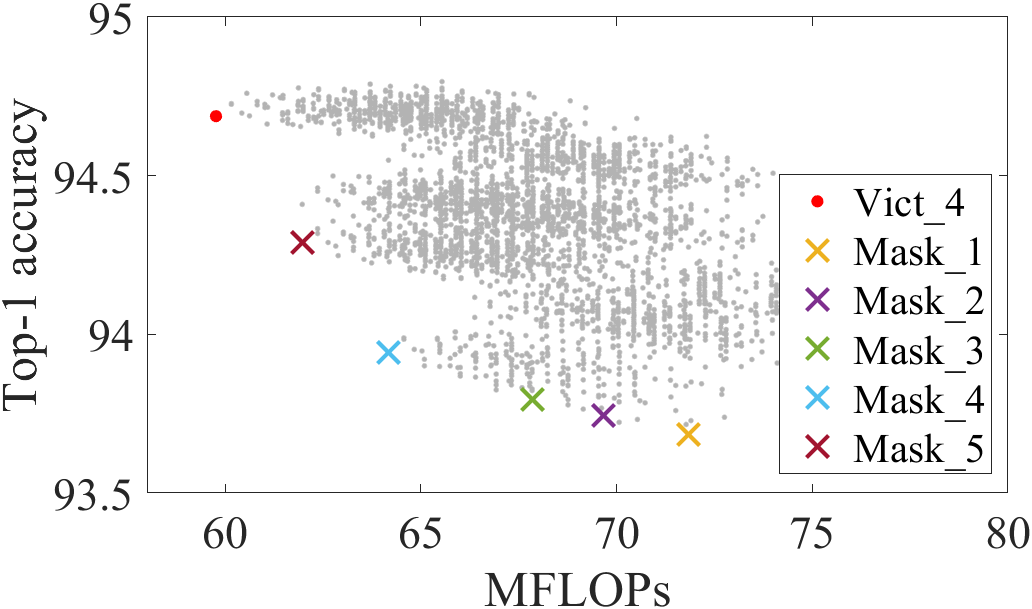}
         \caption{Masks of Vict$\_$4.}
         \label{fig:victim301_mask4}
     \end{subfigure}
        \caption{The optimal mask with different FLOPs overhead for victims in NB-301. }
        \label{fig:victim301_mask}
\end{figure*}

Next, we search for the optimal masks for victims in NB-101 and NB-301 with different FLOPs constraints. The results of NB-101 are shown in Fig. \ref{fig:victim101_mask}, where the optimal masks for Vict\_1 and Vict\_3 are unique regardless of the FLOPs constraints. The reason is that these two masks only adopt the shortcut adding strategy, which is effective for architecture protection but introduces no FLOPs overhead. For Vict\_2, the results show that a mask with lower accuracy can be found if FLOPs overhead increases.  For each victim in NB-301, the optimal mask would be different, as shown in Fig. \ref{fig:victim301_mask}, depending on the given FLOPs budget.

\begin{table*}[htbp]
\centering
\caption{Ablation studies of different obfuscation strategies.}
\label{tab:ablation}
\renewcommand\tabcolsep{12pt} 
\begin{tabular}{cccccccccc}
\toprule
\multirow{2}{*}{} &
  \multicolumn{1}{c}{\multirow{2}{*}{\ Victim \ }} &
  \multicolumn{2}{c}{Original} &
  \multicolumn{2}{c}{Scaling-up} &
  \multicolumn{2}{c}{OP-change} &
  \multicolumn{2}{c}{CN-adding} \\ \cmidrule(r){3-4} \cmidrule(r){5-6} \cmidrule(r){7-8}\cmidrule(r){9-10}
 &
  \multicolumn{1}{c}{} &
  \multicolumn{1}{c}{Acc} &
  \multicolumn{1}{c}{MFLOPs} &
  \multicolumn{1}{c}{Acc}  &
  \multicolumn{1}{c}{MFLOPs} &
  \multicolumn{1}{c}{Acc}  &
  \multicolumn{1}{c}{MFLOPs} &
  \multicolumn{1}{c}{Acc}  &
  \multicolumn{1}{c}{MFLOPs} \\ \hline
\multirow{3}{*}{NB-101} &
  Vict\_1 &
  84.61 &
  37.70 &
  83.47 &
  47.49 &
  - &
  - &
  \textbf{82.65} &
  \textbf{37.70}\\
 &
  Vict\_2 &
  89.63 &
  128.85 &
  88.25 &
  146.29 &
  -&
  -&
  \textbf{88.18}&
  \textbf{128.85}\\
 &
  Vict\_3 &
  93.67 &
  446.62 &
  92.95 &
  456.40 &
  - &
  -&
  \textbf{92.94}&
  \textbf{446.62}\\ \hline
\multirow{4}{*}{NB-301} &
  Vict\_1 &
  92.86 &
  26.95 &
  92.87 &
  27.86 &
  \textbf{91.75} &
  \textbf{39.54} &
  - &
  -\\
 &
  Vict\_2 &
  93.81 &
  33.40 &
  93.79 &
  35.43 &
  \textbf{92.52} &
  \textbf{45.89} &
  - &
  -\\
 &
  Vict\_3 &
  94.46 &
  43.24 &
  94.41 &
  44.16 &
  \textbf{93.23} &
  \textbf{55.72} &
  - &
  - \\
 &
  Vict\_4 &
  94.69 &
  59.76 &
  94.65 &
  63.43 &
  \textbf{93.84} &
  \textbf{68.17} &
  - &
  - \\ \bottomrule
\end{tabular}
\end{table*}
\section{Discussion}
Although the obfuscation performance of scaling-up is not promising for victims in AlphaNet (Table \ref{tab:res}), we note that it likely results from the sophisticated architecture of the super-net of AlphaNet, e.g., including ResNet-like structures, and the jointly optimized weights for both smaller and larger DNNs. To further evaluate the performance of scaling-up, as well as the other two obfuscation strategies, we include the ablation studies on NB-101 and NB-301, for they involve all proposed strategies.  

As shown in Table \ref{tab:ablation}, only adopting connection-adding achieves a higher accuracy drop with lower FLOPs overhead than the scaling-up, although the latter still show up to 1.4\% accuracy drop. For victims in NB-301, the results indicate that the operation-change strategies make main contributions to the best obfuscation performance (Table \ref{tab:res}) compared to the scaling-up strategies. However, both cases demonstrate that by combining the scaling-up strategies with others, the best obfuscation performance can still get improved.

Overall, each strategy has different obfuscation performance, which is also influenced by the victim architecture and its search space. All of them will bring extra FLOPs overhead except the shortcut-adding strategy. For the discussed victim models, the operation-change strategies and connection-adding strategies perform better than the scaling-up strategies, which can be explained by the different optimization difficulties they might bring. Specifically, if the operation-change strategies are applied, it is challenging for attackers to train a Conv layer to make it function as average pooling or identity mapping, and it is also hard to undo the effect of shortcut adding and layer branch adding during training. However, the effect of scaling-up obfuscation would be limited with the sophisticated training strategies, which again demonstrates that the method  used in some existing works, i.e., only enlarging the architectural difference  by scaling-up DNNs, is far from enough for the architectural obfuscation.

\section{Conclusion}
This work presents \ourtit, a NAS-based algorithmic obfuscation approach, to mitigate malicious DNN architecture extractions. 
To prevent attackers from training a competitive substitute model, \ourtit\ minimizes the model accuracy of the extracted architecture while still preserving the original inference accuracy of the victim models. As a generic defense approach, \ourtit\ includes seven function-preserving obfuscation strategies that could increase the optimization difficulties. Leveraging the evolutionary search algorithm, this approach can find the best combination of obfuscation strategies for a victim model. Overall, \ourtit\ can achieve 2.6\% inference accuracy degradation to attackers with only 0.14$\times$ FLOPs overhead, which is 4.7\% better than the SOTA work.

\bibliographystyle{ACM-Reference-Format}

\bibliography{ref.bib}


\begin{thebibliography}{32}


\ifx \showCODEN    \undefined \def \showCODEN     #1{\unskip}     \fi
\ifx \showDOI      \undefined \def \showDOI       #1{#1}\fi
\ifx \showISBNx    \undefined \def \showISBNx     #1{\unskip}     \fi
\ifx \showISBNxiii \undefined \def \showISBNxiii  #1{\unskip}     \fi
\ifx \showISSN     \undefined \def \showISSN      #1{\unskip}     \fi
\ifx \showLCCN     \undefined \def \showLCCN      #1{\unskip}     \fi
\ifx \shownote     \undefined \def \shownote      #1{#1}          \fi
\ifx \showarticletitle \undefined \def \showarticletitle #1{#1}   \fi
\ifx \showURL      \undefined \def \showURL       {\relax}        \fi
\providecommand\bibfield[2]{#2}
\providecommand\bibinfo[2]{#2}
\providecommand\natexlab[1]{#1}
\providecommand\showeprint[2][]{arXiv:#2}

\bibitem[Batina et~al\mbox{.}(2019)]%
        {batina2019csi}
\bibfield{author}{\bibinfo{person}{Lejla Batina}, \bibinfo{person}{Shivam
  Bhasin}, \bibinfo{person}{Dirmanto Jap}, {and} \bibinfo{person}{Stjepan
  Picek}.} \bibinfo{year}{2019}\natexlab{}.
\newblock \showarticletitle{{CSI} {NN:} Reverse engineering of neural network
  architectures through electromagnetic side channel}. In
  \bibinfo{booktitle}{\emph{28th USENIX Security Symposium (USENIX Security
  19)}}. \bibinfo{pages}{515--532}.
\newblock


\bibitem[Bender et~al\mbox{.}(2018)]%
        {bender2018understanding}
\bibfield{author}{\bibinfo{person}{Gabriel Bender}, \bibinfo{person}{Pieter-Jan
  Kindermans}, \bibinfo{person}{Barret Zoph}, \bibinfo{person}{Vijay
  Vasudevan}, {and} \bibinfo{person}{Quoc Le}.}
  \bibinfo{year}{2018}\natexlab{}.
\newblock \showarticletitle{Understanding and simplifying one-shot architecture
  search}. In \bibinfo{booktitle}{\emph{International Conference on Machine
  Learning (ICML)}}. PMLR, \bibinfo{pages}{550--559}.
\newblock


\bibitem[He et~al\mbox{.}(2016)]%
        {he2016deep}
\bibfield{author}{\bibinfo{person}{Kaiming He}, \bibinfo{person}{Xiangyu
  Zhang}, \bibinfo{person}{Shaoqing Ren}, {and} \bibinfo{person}{Jian Sun}.}
  \bibinfo{year}{2016}\natexlab{}.
\newblock \showarticletitle{Deep residual learning for image recognition}. In
  \bibinfo{booktitle}{\emph{Proceedings of the IEEE/CVF Conference on Computer
  Vision and Pattern Recognition (CVPR)}}. \bibinfo{pages}{770--778}.
\newblock


\bibitem[He et~al\mbox{.}(2021)]%
        {he2021automl}
\bibfield{author}{\bibinfo{person}{Xin He}, \bibinfo{person}{Kaiyong Zhao},
  {and} \bibinfo{person}{Xiaowen Chu}.} \bibinfo{year}{2021}\natexlab{}.
\newblock \showarticletitle{AutoML: A survey of the state-of-the-art}.
\newblock \bibinfo{journal}{\emph{Knowledge-Based Systems}}
  \bibinfo{volume}{212} (\bibinfo{year}{2021}), \bibinfo{pages}{106622}.
\newblock


\bibitem[Hu et~al\mbox{.}(2020)]%
        {hu2020deepsniffer}
\bibfield{author}{\bibinfo{person}{Xing Hu}, \bibinfo{person}{Ling Liang},
  \bibinfo{person}{Shuangchen Li}, \bibinfo{person}{Lei Deng},
  \bibinfo{person}{Pengfei Zuo}, \bibinfo{person}{Yu Ji},
  \bibinfo{person}{Xinfeng Xie}, \bibinfo{person}{Yufei Ding},
  \bibinfo{person}{Chang Liu}, \bibinfo{person}{Timothy Sherwood},
  {et~al\mbox{.}}} \bibinfo{year}{2020}\natexlab{}.
\newblock \showarticletitle{Deepsniffer: A dnn model extraction framework based
  on learning architectural hints}. In \bibinfo{booktitle}{\emph{Proceedings of
  the Twenty-Fifth International Conference on Architectural Support for
  Programming Languages and Operating Systems}}. \bibinfo{pages}{385--399}.
\newblock


\bibitem[Hua et~al\mbox{.}(2018)]%
        {hua2018reverse}
\bibfield{author}{\bibinfo{person}{Weizhe Hua}, \bibinfo{person}{Zhiru Zhang},
  {and} \bibinfo{person}{G~Edward Suh}.} \bibinfo{year}{2018}\natexlab{}.
\newblock \showarticletitle{Reverse engineering convolutional neural networks
  through side-channel information leaks}. In \bibinfo{booktitle}{\emph{2018
  55th ACM/ESDA/IEEE Design Automation Conference (DAC)}}. IEEE,
  \bibinfo{pages}{1--6}.
\newblock


\bibitem[Ioffe and Szegedy(2015)]%
        {ioffe2015batch}
\bibfield{author}{\bibinfo{person}{Sergey Ioffe} {and}
  \bibinfo{person}{Christian Szegedy}.} \bibinfo{year}{2015}\natexlab{}.
\newblock \showarticletitle{Batch normalization: Accelerating deep network
  training by reducing internal covariate shift}. In
  \bibinfo{booktitle}{\emph{International conference on machine learning
  (ICML)}}. PMLR, \bibinfo{pages}{448--456}.
\newblock


\bibitem[Krizhevsky et~al\mbox{.}(2009)]%
        {cifar10_data}
\bibfield{author}{\bibinfo{person}{Alex Krizhevsky}, \bibinfo{person}{Vinod
  Nair}, {and} \bibinfo{person}{Geoffrey Hinton}.}
  \bibinfo{year}{2009}\natexlab{}.
\newblock \showarticletitle{CIFAR-10 (Canadian Institute for Advanced
  Research)}.
\newblock  (\bibinfo{year}{2009}).
\newblock
\urldef\tempurl%
\url{http://www.cs.toronto.edu/~kriz/cifar.html}
\showURL{%
\tempurl}


\bibitem[Li et~al\mbox{.}(2021)]%
        {li2021neurobfuscator}
\bibfield{author}{\bibinfo{person}{Jingtao Li}, \bibinfo{person}{Zhezhi He},
  \bibinfo{person}{Adnan~Siraj Rakin}, \bibinfo{person}{Deliang Fan}, {and}
  \bibinfo{person}{Chaitali Chakrabarti}.} \bibinfo{year}{2021}\natexlab{}.
\newblock \showarticletitle{NeurObfuscator: {A} Full-stack Obfuscation Tool to
  Mitigate Neural Architecture Stealing}. In \bibinfo{booktitle}{\emph{{IEEE}
  International Symposium on Hardware Oriented Security and Trust (HOST)}}.
  \bibinfo{publisher}{{IEEE}}, \bibinfo{pages}{248--258}.
\newblock


\bibitem[Liu et~al\mbox{.}(2018b)]%
        {liu2018progressive}
\bibfield{author}{\bibinfo{person}{Chenxi Liu}, \bibinfo{person}{Barret Zoph},
  \bibinfo{person}{Maxim Neumann}, \bibinfo{person}{Jonathon Shlens},
  \bibinfo{person}{Wei Hua}, \bibinfo{person}{Li-Jia Li}, \bibinfo{person}{Li
  Fei-Fei}, \bibinfo{person}{Alan Yuille}, \bibinfo{person}{Jonathan Huang},
  {and} \bibinfo{person}{Kevin Murphy}.} \bibinfo{year}{2018}\natexlab{b}.
\newblock \showarticletitle{Progressive neural architecture search}. In
  \bibinfo{booktitle}{\emph{Proceedings of the European conference on computer
  vision (ECCV)}}. \bibinfo{pages}{19--34}.
\newblock


\bibitem[Liu et~al\mbox{.}(2018a)]%
        {liu2017hierarchical}
\bibfield{author}{\bibinfo{person}{Hanxiao Liu}, \bibinfo{person}{Karen
  Simonyan}, \bibinfo{person}{Oriol Vinyals}, \bibinfo{person}{Chrisantha
  Fernando}, {and} \bibinfo{person}{Koray Kavukcuoglu}.}
  \bibinfo{year}{2018}\natexlab{a}.
\newblock \showarticletitle{Hierarchical representations for efficient
  architecture search}.
\newblock \bibinfo{journal}{\emph{6th International Conference on Learning
  Representations (ICLR)}} (\bibinfo{year}{2018}).
\newblock


\bibitem[Liu et~al\mbox{.}(2019b)]%
        {liu2018darts}
\bibfield{author}{\bibinfo{person}{Hanxiao Liu}, \bibinfo{person}{Karen
  Simonyan}, {and} \bibinfo{person}{Yiming Yang}.}
  \bibinfo{year}{2019}\natexlab{b}.
\newblock \showarticletitle{Darts: Differentiable architecture search}.
\newblock \bibinfo{journal}{\emph{7th International Conference on Learning
  Representations (ICLR)}} (\bibinfo{year}{2019}).
\newblock


\bibitem[Liu et~al\mbox{.}(2019a)]%
        {liu2019mitigating}
\bibfield{author}{\bibinfo{person}{Yuntao Liu}, \bibinfo{person}{Dana
  Dachman-Soled}, {and} \bibinfo{person}{Ankur Srivastava}.}
  \bibinfo{year}{2019}\natexlab{a}.
\newblock \showarticletitle{Mitigating reverse engineering attacks on deep
  neural networks}. In \bibinfo{booktitle}{\emph{2019 IEEE Computer Society
  Annual Symposium on VLSI (ISVLSI)}}. IEEE, \bibinfo{pages}{657--662}.
\newblock


\bibitem[Luo et~al\mbox{.}(2022)]%
        {luo2022nnrearch}
\bibfield{author}{\bibinfo{person}{Yukui Luo}, \bibinfo{person}{Shijin Duan},
  \bibinfo{person}{Cheng Gongye}, \bibinfo{person}{Yunsi Fei}, {and}
  \bibinfo{person}{Xiaolin Xu}.} \bibinfo{year}{2022}\natexlab{}.
\newblock \showarticletitle{NNReArch: A Tensor Program Scheduling Framework
  Against Neural Network Architecture Reverse Engineering}. In
  \bibinfo{booktitle}{\emph{2022 IEEE 30th Annual International Symposium on
  Field-Programmable Custom Computing Machines (FCCM)}}. IEEE,
  \bibinfo{pages}{1--9}.
\newblock


\bibitem[Real et~al\mbox{.}(2019)]%
        {real2019regularized}
\bibfield{author}{\bibinfo{person}{Esteban Real}, \bibinfo{person}{Alok
  Aggarwal}, \bibinfo{person}{Yanping Huang}, {and} \bibinfo{person}{Quoc~V
  Le}.} \bibinfo{year}{2019}\natexlab{}.
\newblock \showarticletitle{Regularized evolution for image classifier
  architecture search}. In \bibinfo{booktitle}{\emph{The Thirty-Third {AAAI}
  Conference on Artificial Intelligence}}, Vol.~\bibinfo{volume}{33}.
  \bibinfo{pages}{4780--4789}.
\newblock


\bibitem[Sandler et~al\mbox{.}(2018)]%
        {sandler2018mobilenetv2}
\bibfield{author}{\bibinfo{person}{Mark Sandler}, \bibinfo{person}{Andrew
  Howard}, \bibinfo{person}{Menglong Zhu}, \bibinfo{person}{Andrey Zhmoginov},
  {and} \bibinfo{person}{Liang-Chieh Chen}.} \bibinfo{year}{2018}\natexlab{}.
\newblock \showarticletitle{Mobilenetv2: Inverted residuals and linear
  bottlenecks}. In \bibinfo{booktitle}{\emph{Proceedings of the IEEE/CVF
  Conference on Computer Vision and Pattern Recognition (CVPR)}}.
  \bibinfo{pages}{4510--4520}.
\newblock


\bibitem[Siems et~al\mbox{.}(2020)]%
        {siems2020bench}
\bibfield{author}{\bibinfo{person}{Julien Siems}, \bibinfo{person}{Lucas
  Zimmer}, \bibinfo{person}{Arber Zela}, \bibinfo{person}{Jovita Lukasik},
  \bibinfo{person}{Margret Keuper}, {and} \bibinfo{person}{Frank Hutter}.}
  \bibinfo{year}{2020}\natexlab{}.
\newblock \showarticletitle{Nas-bench-301 and the case for surrogate benchmarks
  for neural architecture search}.
\newblock \bibinfo{journal}{\emph{arXiv preprint arXiv:2008.09777}}
  (\bibinfo{year}{2020}).
\newblock


\bibitem[Simonyan and Zisserman(2015)]%
        {simonyan2014very}
\bibfield{author}{\bibinfo{person}{Karen Simonyan} {and}
  \bibinfo{person}{Andrew Zisserman}.} \bibinfo{year}{2015}\natexlab{}.
\newblock \showarticletitle{Very deep convolutional networks for large-scale
  image recognition}.
\newblock \bibinfo{journal}{\emph{3rd International Conference on Learning
  Representations (ICLR)}} (\bibinfo{year}{2015}).
\newblock


\bibitem[Srivastava et~al\mbox{.}(2015)]%
        {srivastava2015training}
\bibfield{author}{\bibinfo{person}{Rupesh~K Srivastava}, \bibinfo{person}{Klaus
  Greff}, {and} \bibinfo{person}{J{\"u}rgen Schmidhuber}.}
  \bibinfo{year}{2015}\natexlab{}.
\newblock \showarticletitle{Training very deep networks}.
\newblock \bibinfo{journal}{\emph{Advances in neural information processing
  systems}}  \bibinfo{volume}{28} (\bibinfo{year}{2015}).
\newblock


\bibitem[Suganuma et~al\mbox{.}(2017)]%
        {suganuma2017genetic}
\bibfield{author}{\bibinfo{person}{Masanori Suganuma},
  \bibinfo{person}{Shinichi Shirakawa}, {and} \bibinfo{person}{Tomoharu
  Nagao}.} \bibinfo{year}{2017}\natexlab{}.
\newblock \showarticletitle{A genetic programming approach to designing
  convolutional neural network architectures}. In
  \bibinfo{booktitle}{\emph{Proceedings of the genetic and evolutionary
  computation conference}}. \bibinfo{pages}{497--504}.
\newblock


\bibitem[Szegedy et~al\mbox{.}(2017)]%
        {szegedy2017inception}
\bibfield{author}{\bibinfo{person}{Christian Szegedy}, \bibinfo{person}{Sergey
  Ioffe}, \bibinfo{person}{Vincent Vanhoucke}, {and}
  \bibinfo{person}{Alexander~A Alemi}.} \bibinfo{year}{2017}\natexlab{}.
\newblock \showarticletitle{Inception-v4, inception-resnet and the impact of
  residual connections on learning}. In \bibinfo{booktitle}{\emph{Thirty-first
  AAAI conference on artificial intelligence}}.
\newblock


\bibitem[Tan et~al\mbox{.}(2019)]%
        {tan2019mnasnet}
\bibfield{author}{\bibinfo{person}{Mingxing Tan}, \bibinfo{person}{Bo Chen},
  \bibinfo{person}{Ruoming Pang}, \bibinfo{person}{Vijay Vasudevan},
  \bibinfo{person}{Mark Sandler}, \bibinfo{person}{Andrew Howard}, {and}
  \bibinfo{person}{Quoc~V Le}.} \bibinfo{year}{2019}\natexlab{}.
\newblock \showarticletitle{Mnasnet: Platform-aware neural architecture search
  for mobile}. In \bibinfo{booktitle}{\emph{Proceedings of the IEEE/CVF
  Conference on Computer Vision and Pattern Recognition (CVPR)}}.
  \bibinfo{pages}{2820--2828}.
\newblock


\bibitem[Wang et~al\mbox{.}(2021a)]%
        {wang2021alphanet}
\bibfield{author}{\bibinfo{person}{Dilin Wang}, \bibinfo{person}{Chengyue
  Gong}, \bibinfo{person}{Meng Li}, \bibinfo{person}{Qiang Liu}, {and}
  \bibinfo{person}{Vikas Chandra}.} \bibinfo{year}{2021}\natexlab{a}.
\newblock \showarticletitle{AlphaNet: Improved Training of Supernets with
  Alpha-Divergence}. In \bibinfo{booktitle}{\emph{International Conference on
  Machine Learning (ICML)}}. PMLR, \bibinfo{pages}{10760--10771}.
\newblock


\bibitem[Wang et~al\mbox{.}(2021b)]%
        {wang2021attentivenas}
\bibfield{author}{\bibinfo{person}{Dilin Wang}, \bibinfo{person}{Meng Li},
  \bibinfo{person}{Chengyue Gong}, {and} \bibinfo{person}{Vikas Chandra}.}
  \bibinfo{year}{2021}\natexlab{b}.
\newblock \showarticletitle{Attentivenas: Improving neural architecture search
  via attentive sampling}. In \bibinfo{booktitle}{\emph{Proceedings of the
  IEEE/CVF Conference on Computer Vision and Pattern Recognition (CVPR)}}.
  \bibinfo{pages}{6418--6427}.
\newblock


\bibitem[Wang et~al\mbox{.}(2019)]%
        {wang2019npufort}
\bibfield{author}{\bibinfo{person}{Xingbin Wang}, \bibinfo{person}{Rui Hou},
  \bibinfo{person}{Yifan Zhu}, \bibinfo{person}{Jun Zhang}, {and}
  \bibinfo{person}{Dan Meng}.} \bibinfo{year}{2019}\natexlab{}.
\newblock \showarticletitle{NPUFort: A secure architecture of DNN accelerator
  against model inversion attack}. In \bibinfo{booktitle}{\emph{Proceedings of
  the 16th ACM International Conference on Computing Frontiers}}.
  \bibinfo{pages}{190--196}.
\newblock


\bibitem[Wei et~al\mbox{.}(2022)]%
        {wei2022npenas}
\bibfield{author}{\bibinfo{person}{Chen Wei}, \bibinfo{person}{Chuang Niu},
  \bibinfo{person}{Yiping Tang}, \bibinfo{person}{Yue Wang},
  \bibinfo{person}{Haihong Hu}, {and} \bibinfo{person}{Jimin Liang}.}
  \bibinfo{year}{2022}\natexlab{}.
\newblock \showarticletitle{Npenas: Neural predictor guided evolution for
  neural architecture search}.
\newblock \bibinfo{journal}{\emph{IEEE Transactions on Neural Networks and
  Learning Systems}} (\bibinfo{year}{2022}).
\newblock


\bibitem[Wei et~al\mbox{.}(2020)]%
        {wei2020leaky}
\bibfield{author}{\bibinfo{person}{Junyi Wei}, \bibinfo{person}{Yicheng Zhang},
  \bibinfo{person}{Zhe Zhou}, \bibinfo{person}{Zhou Li}, {and}
  \bibinfo{person}{Mohammad~Abdullah Al~Faruque}.}
  \bibinfo{year}{2020}\natexlab{}.
\newblock \showarticletitle{Leaky DNN: Stealing deep-learning model secret with
  GPU context-switching side-channel}. In \bibinfo{booktitle}{\emph{2020 50th
  Annual IEEE/IFIP International Conference on Dependable Systems and Networks
  (DSN)}}. IEEE, \bibinfo{pages}{125--137}.
\newblock


\bibitem[Wen et~al\mbox{.}(2020)]%
        {wen2020neural}
\bibfield{author}{\bibinfo{person}{Wei Wen}, \bibinfo{person}{Hanxiao Liu},
  \bibinfo{person}{Yiran Chen}, \bibinfo{person}{Hai Li},
  \bibinfo{person}{Gabriel Bender}, {and} \bibinfo{person}{Pieter-Jan
  Kindermans}.} \bibinfo{year}{2020}\natexlab{}.
\newblock \showarticletitle{Neural predictor for neural architecture search}.
  In \bibinfo{booktitle}{\emph{Proceedings of the European conference on
  computer vision (ECCV)}}. \bibinfo{pages}{660--676}.
\newblock


\bibitem[Yan et~al\mbox{.}(2020)]%
        {yan2020cache}
\bibfield{author}{\bibinfo{person}{Mengjia Yan}, \bibinfo{person}{Christopher~W
  Fletcher}, {and} \bibinfo{person}{Josep Torrellas}.}
  \bibinfo{year}{2020}\natexlab{}.
\newblock \showarticletitle{Cache telepathy: Leveraging shared resource attacks
  to learn DNN architectures}. In \bibinfo{booktitle}{\emph{29th USENIX
  Security Symposium (USENIX Security 20)}}. \bibinfo{pages}{2003--2020}.
\newblock


\bibitem[Ying et~al\mbox{.}(2019)]%
        {ying2019bench}
\bibfield{author}{\bibinfo{person}{Chris Ying}, \bibinfo{person}{Aaron Klein},
  \bibinfo{person}{Eric Christiansen}, \bibinfo{person}{Esteban Real},
  \bibinfo{person}{Kevin Murphy}, {and} \bibinfo{person}{Frank Hutter}.}
  \bibinfo{year}{2019}\natexlab{}.
\newblock \showarticletitle{Nas-bench-101: Towards reproducible neural
  architecture search}. In \bibinfo{booktitle}{\emph{International Conference
  on Machine Learning (ICML)}}. PMLR, \bibinfo{pages}{7105--7114}.
\newblock


\bibitem[Yu et~al\mbox{.}(2020)]%
        {yu2020deepem}
\bibfield{author}{\bibinfo{person}{Honggang Yu}, \bibinfo{person}{Haocheng Ma},
  \bibinfo{person}{Kaichen Yang}, \bibinfo{person}{Yiqiang Zhao}, {and}
  \bibinfo{person}{Yier Jin}.} \bibinfo{year}{2020}\natexlab{}.
\newblock \showarticletitle{Deepem: Deep neural networks model recovery through
  em side-channel information leakage}. In \bibinfo{booktitle}{\emph{2020 IEEE
  International Symposium on Hardware Oriented Security and Trust (HOST)}}.
  IEEE, \bibinfo{pages}{209--218}.
\newblock


\bibitem[Zhu et~al\mbox{.}(2021)]%
        {zhu2021hermes}
\bibfield{author}{\bibinfo{person}{Yuankun Zhu}, \bibinfo{person}{Yueqiang
  Cheng}, \bibinfo{person}{Husheng Zhou}, {and} \bibinfo{person}{Yantao Lu}.}
  \bibinfo{year}{2021}\natexlab{}.
\newblock \showarticletitle{Hermes attack: Steal DNN models with lossless
  inference accuracy}. In \bibinfo{booktitle}{\emph{30th USENIX Security
  Symposium (USENIX Security 21)}}. \bibinfo{pages}{1973--1988}.
\newblock


\end{thebibliography}

\end{document}